\documentstyle[12pt]{article}
\begin{document}

\newcommand\Durkop{1}
\newcommand\ElmersA{2}
\newcommand\Elmersb{3}
\newcommand\ElmersB{4}
\newcommand\Sandera{5}
\newcommand\Sanderb{6}
\newcommand\Skomski{7}
\newcommand\Bethge{8}
\newcommand\Elmersa{9}
\newcommand\Elmersc{10}
\newcommand\Suen{11}
\newcommand\Kolesik{12}
\newcommand\Pokhil{13}
\newcommand\Gadetsky{14}
\newcommand\Kirilyuk{15}
\newcommand\Skomskib{16}
\newcommand\Rikvold{17}
\newcommand\Kolesikb{18}

\begin{center}
{\bf 
MONTE CARLO SIMULATION OF MAGNETIZATION REVERSAL \\ 
VIA DOMAIN-WALL MOTION IN Fe SESQUILAYERS ON W(110)}
\end{center}
M. KOLESIK$^{1,2}$,
M. A. NOVOTNY$^{1,3}$,
and
PER ARNE RIKVOLD$^{1,4}$ 
\\ 
$^1$Supercomputer Computations Research Institute, \\
\hspace*{0.2truecm}Florida State University, 
                   Tallahassee, Florida 32306-4130 \\
$^2$Institute of Physics, 
    Slovak Academy of Sciences, \\
\hspace*{0.2truecm}D\' ubravsk\' a cesta 9, 
                   84228 Bratislava, Slovak Republic \\
$^3$Department of Electrical Engineering, 2525 Pottsdamer Street, \\   
\hspace*{0.2truecm}Florida A\&M University--Florida State University,
                   Tallahassee, Florida 32310-6046 \\
$^4$Center for Materials Research and Technology and Department of Physics,\\ 
\hspace*{0.2truecm}Florida State University, 
                   Tallahassee, Florida 32306-4350

\vspace{0.3truecm}
\noindent ABSTRACT
\vspace{0.4truecm}

Iron sesquilayers are ultrathin films with coverages between
one and two atomic monolayers. They consist of an almost
defect-free monolayer with compact islands of a second atomic
layer on top. This variation of the  film thickness results in a 
strong interaction
between domain walls and the island structure. It makes these
systems an ideal laboratory to study the dynamics of domain walls
driven by weak external fields. We present computer simulations
which provide insight into the role of the thermally activated 
nucleation processes by which a driven domain wall overcomes
the obstacles created by the islands.

\vspace{0.4truecm}
\noindent INTRODUCTION 
\vspace{0.4truecm}

Iron sesquilayers with coverages between one and two atomic monolayers
grown at room temperature on the (110) surface of tungsten are extremely 
interesting systems, both from the experimental and 
theoretical point of view. 
There have been continuous experimental efforts [\Durkop --\Suen ] 
to understand the physics of these films, and surprising results still emerge 
[\Durkop , \ElmersA ].
Below their critical temperatures, these films are uniaxial in-plane
ferromagnets which may belong to the Ising universality class 
[\ElmersA , \Elmersb , \ElmersB ].
As a function of coverage, their coercivity exhibits a pronounced maximum 
located around 1.4 atomic monolayers [\Sandera , \Sanderb , \Skomski ].
Recently, a spin reorientation transition was observed in freshly prepared
samples at temperatures above the Curie point of a monolayer [\Durkop ].
The origin of these unusual magnetic  properties lies in the film morphology.
These ultrathin films consist of a nearly perfect atomic monolayer which
supports compact islands of the second monolayer [\Bethge ], as is illustrated
in Fig.~1. The island structure results in a varying film thickness 
which causes a domain wall to prefer certain configurations. 
That the morphology can be controlled by the film thickness
and that the surface can be precisely mapped by STM make
sesquilayers ideal systems in which to study the role of disorder in  
domain-wall propagation.

Recently, we developed a computer model for iron sesquilayers
and utilized it to simulate  domain-wall motion in a magnetic field [\Kolesik ].
This enabled us to describe the dependence of the coercivity on the
film thickness, the  temperature, and the frequency of the field. 
The success of the model motivated us
to look more closely at the processes involved in  domain-wall
motion. This is the subject of the present work. We address
questions concerning the functional dependence of the domain-wall velocity
on the driving field. There are different models used in the
literature which differ significantly [\Pokhil --\Skomskib], and  
there is no generally accepted opinion on the field dependence of 
the activation volumes and energies involved in domain-wall motion
driven by weak fields. 
Therefore, we use Monte Carlo simulations to obtain information about 
the interaction between a domain wall and the pinning environment of rough films. 
We concentrate on quantities which have not yet been directly observed in experiments.
In simulations it is possible to identify the metastable domain-wall 
configurations, the energy barriers which separate them, the
activation volumes controlling the thermally activated domain-wall motion,
and the  Barkhausen volumes. 
Our results suggest that quantities such as the activation volume or energy must 
be used with  great care because what is observed in experiments are effective quantities 
reflecting a complex interplay between disorder, temperature, driving field and, 
possibly, the history of the sample.

\vspace{ -10.5 truecm}
{\Large
\hspace{6.5 truecm} (a) \hspace{7 truecm} (b)
}
\vspace{  10 truecm}

\begin{figure}[t]
\vspace{8.2 truecm}
 \includegraphics{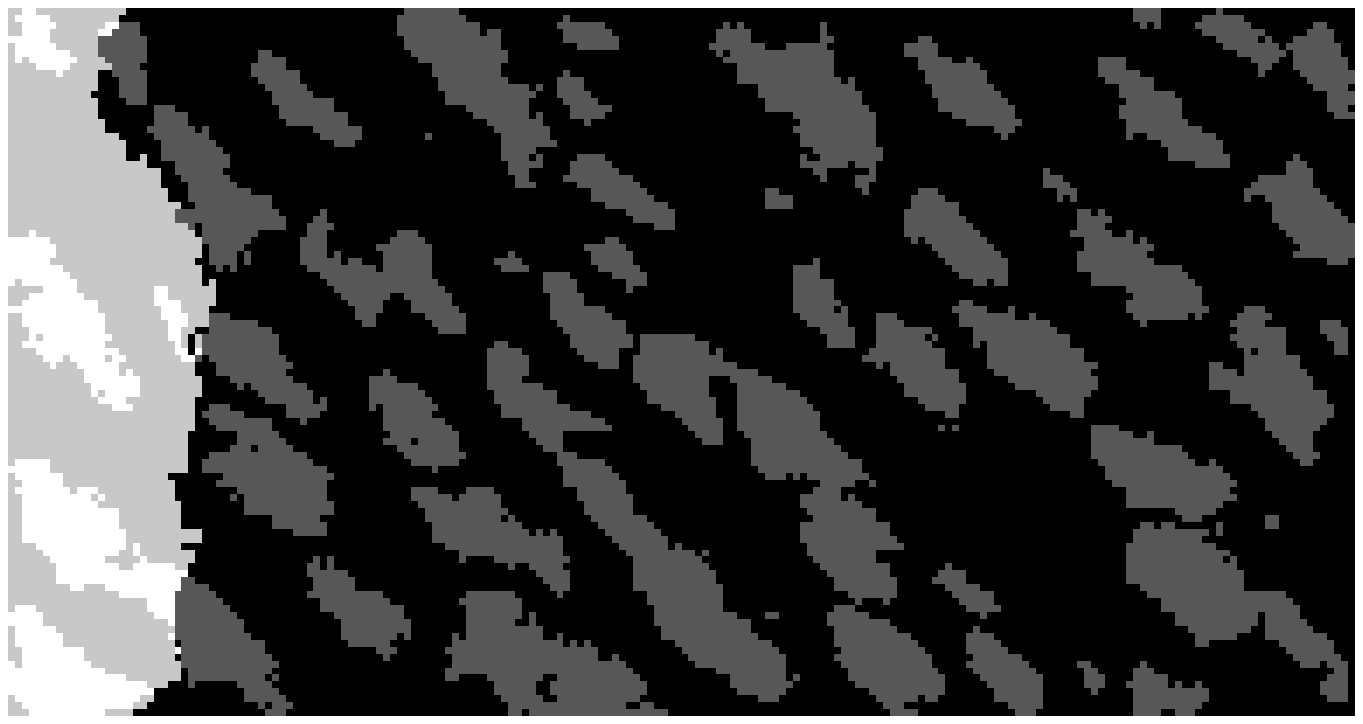}
 \includegraphics{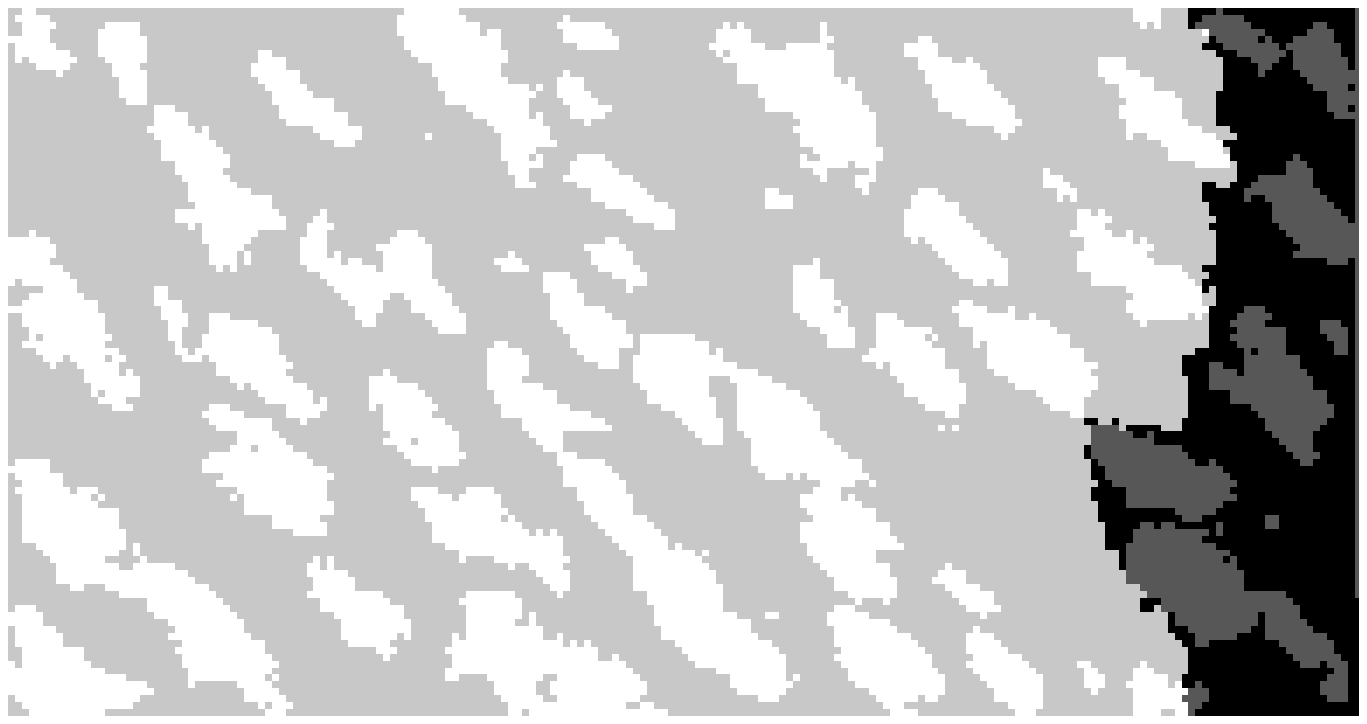}
 \includegraphics{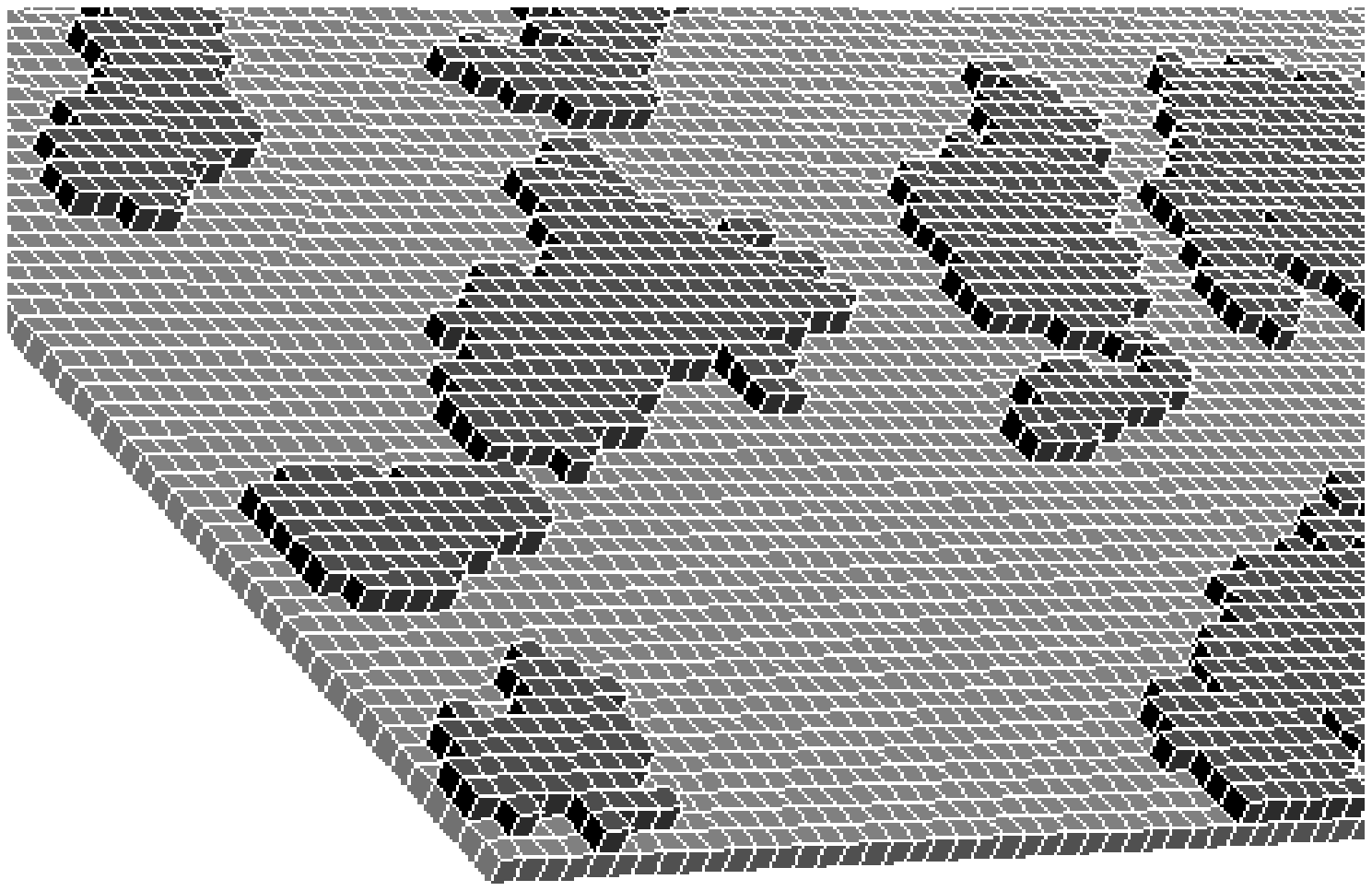}
\caption[]{
(a): 
The sesquilayer morphology and our computer simulation setup.
The patches are islands of the second atomic monolayer on top
of a nearly perfect first monolayer. The coverage of this film
is 1.26 atomic monolayers.
The snapshots show the 1170$\times$612 \AA$^2$ system 
shortly after starting the simulation with a narrow
strip of the stable phase (light) on the left, and just before stopping
the measurements after the domain wall has traveled the length of the sample.
(b): 
The lattice discretization. This is a section of a computational lattice created
from data obtained by digitizing STM pictures published in Ref. [\Bethge ].
}
\end{figure}

\vspace{0.4truecm}
\noindent COMPUTER MODEL OF AN Fe SESQUILAYER FILM
\vspace{0.4truecm}

In this section, we describe our computational model
of iron sesquilayers as developed in Ref.~[\Kolesik ]. 
It is based on the fact that the film is a highly
anisotropic uniaxial ferromagnet [\Elmersc ]. Therefore, the
kinetic Ising model is a reasonable approximation which
can capture the important physics of the system.
The Hamiltonian of the model is that of the usual nearest-neighbor
Ising ferromagnet subject to an external field $H$:
\begin{equation}
{\cal H} = -J \sum_{\langle ij \rangle } s_i s_j - \mu H \sum_{i} s_i \ .
\label{eq:ham}
\end{equation}
The two-state variables $s_i$ describe the two preferred
orientations of the local magnetization and are defined
on a computational lattice which faithfully reflects the
morphology of a real sesquilayer film. Data necessary to
create simulational lattices for various coverages were
obtained by digitizing STM pictures of iron sesquilayers
published by Bethge et al. [\Bethge ]. 
The morphology of the computational lattice is illustrated in Fig.~1.
We have chosen the computational spin $s_i$ to represent an area of a
monolayer of size 6\AA$\times$6\AA. One and two layers of a square lattice 
were used for the mono- and double-layer parts of the film, respectively,
and the spin-spin interaction, $J=8.73$ meV, was fixed such that the
critical temperature of a perfect monolayer (230K [\Elmersb ]) is reproduced.
The size of the computational lattice was 195$\times$102$\times$1(2),
corresponding to 1170$\times$612 \AA$^2$.
To fix the coupling to the external field we use the bulk
magnetic moment for an iron atom. Thus, our computational spins
each carry a magnetic moment of $\mu = 11.43~\mu_B$.
In this work, we only present data for the coverage 1.26 atomic layers at
the temperature of 184 K.

\vspace{0.4truecm}
\noindent SIMULATIONS AND MEASURED QUANTITIES
\vspace{0.4truecm}

In order to obtain detailed information on the dynamics of a domain wall,
we set up a computer experiment with a propagating domain wall driven
by the field. We start with the lattice magnetized as shown in the upper
panel of Fig.~1(a).
A narrow strip along one  short side of the lattice 
is initialized in the stable phase, while the rest
of the system is in the metastable phase. An external field is applied,
and quantities related to the motion of the domain wall are recorded
as the wall propagates. 
Results are averaged over on the order of 100 runs for each value of $H$.
Two important quantities are the average probability
that the volume of the stable phase will grow,  $g(n)$, or shrink, $s(n)$, 
in the next Monte Carlo step.
They are sampled for each value of the 
stable-phase volume 
(measured by the number $n$ of spins in the stable state). Since the
magnetization reversal occurs via domain-wall motion, $n$ translates
into an ``average'' position of the wall.
These growth and shrinkage probabilities
carry information relevant for describing the dynamics of the domain
wall. In particular, they allow us to calculate the average time the domain wall
needs to travel a certain distance, which can be checked against the directly
measured times. The good agreement indicates that the growth and shrinkage 
probabilities can be used reliably to investigate the motion of the wall.

\vspace{0.4truecm}
\noindent DYNAMICS OF THE DOMAIN-WALL MOTION
\vspace{0.4truecm}

The island structure of the second
atomic monolayer creates regions where the domain-wall exchange
energy is about twice as large as in a monolayer.
Thus, there is a complicated free-energy landscape for the domain-wall
configurations. As the wall propagates in this disordered environment,
it prefers to stay in metastable configurations that minimize the
interface free energy. To proceed to the next favorable
position, the domain wall must cross a region where it has an increased
interface free energy. Such regions are usually related to the islands
and represent obstacles or barriers to be overcome before the domain wall
can move forward. The mechanism which allows these obstacles 
to be overcome is provided by thermal fluctuations of the domain wall.
Like in ordinary nucleation phenomena [\Rikvold ], if a fluctuation
into a region where the interfacial free energy is unfavorable
is sufficiently large, the free-energy reduction obtained by further 
increasing the volume of the stable phase just outweighs the cost 
of creating more interface.
The probability for the fluctuation to grow
even further then  becomes larger than the probability to shrink.
Such a fluctuation is called {\em critical}. Typically, the system needs 
many ``attempts'' to create a critical
fluctuation, and it is therefore confined to the metastable 
configurations most of the time. When a thermal ``kick'' is
strong enough, it creates a fluctuation which is supercritical 
and continues to grow. The domain wall then rapidly 
proceeds towards the next metastable configuration, and the whole process 
repeats itself with a new domain-wall shape. Thus, the movement of the wall is jerky,
exhibiting sudden jumps [\Kolesik ]. 
In the experimental literature
the volume of the critical fluctuation is
usually called the activation volume  and
the associated free-energy barrier is  called the activation
energy. These ideas are often used to construct theoretical
models describing the dependence of the domain-wall velocity
on the driving field. It is our goal to explore to what extent
these ideas can be used in the case of sesquilayers. The method we employ 
allows us to measure  the parameters of the critical
fluctuations, including their activation volumes and energies.

\begin{figure}[t]
\vspace{7.5truecm}
 \includegraphics{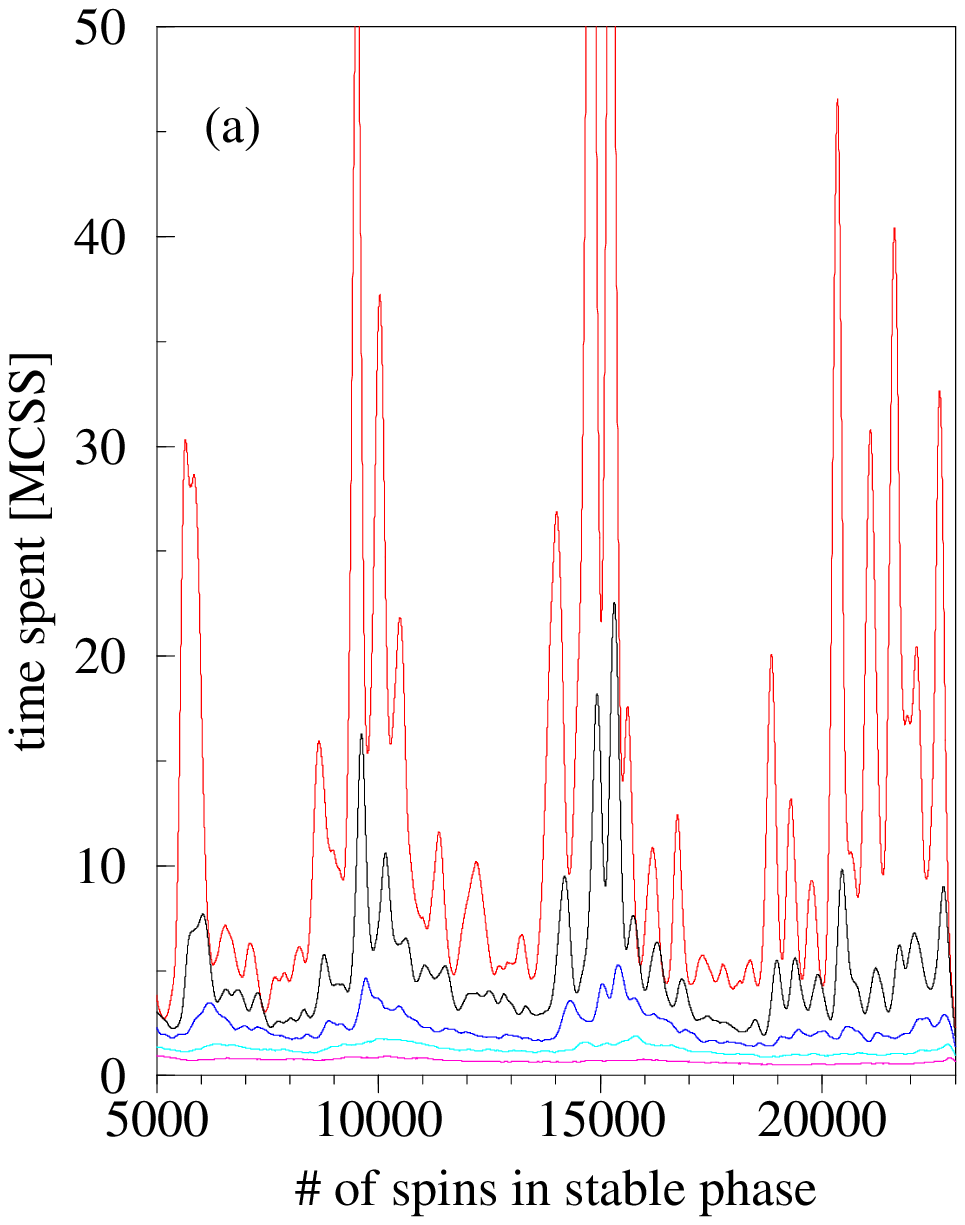}
 \includegraphics{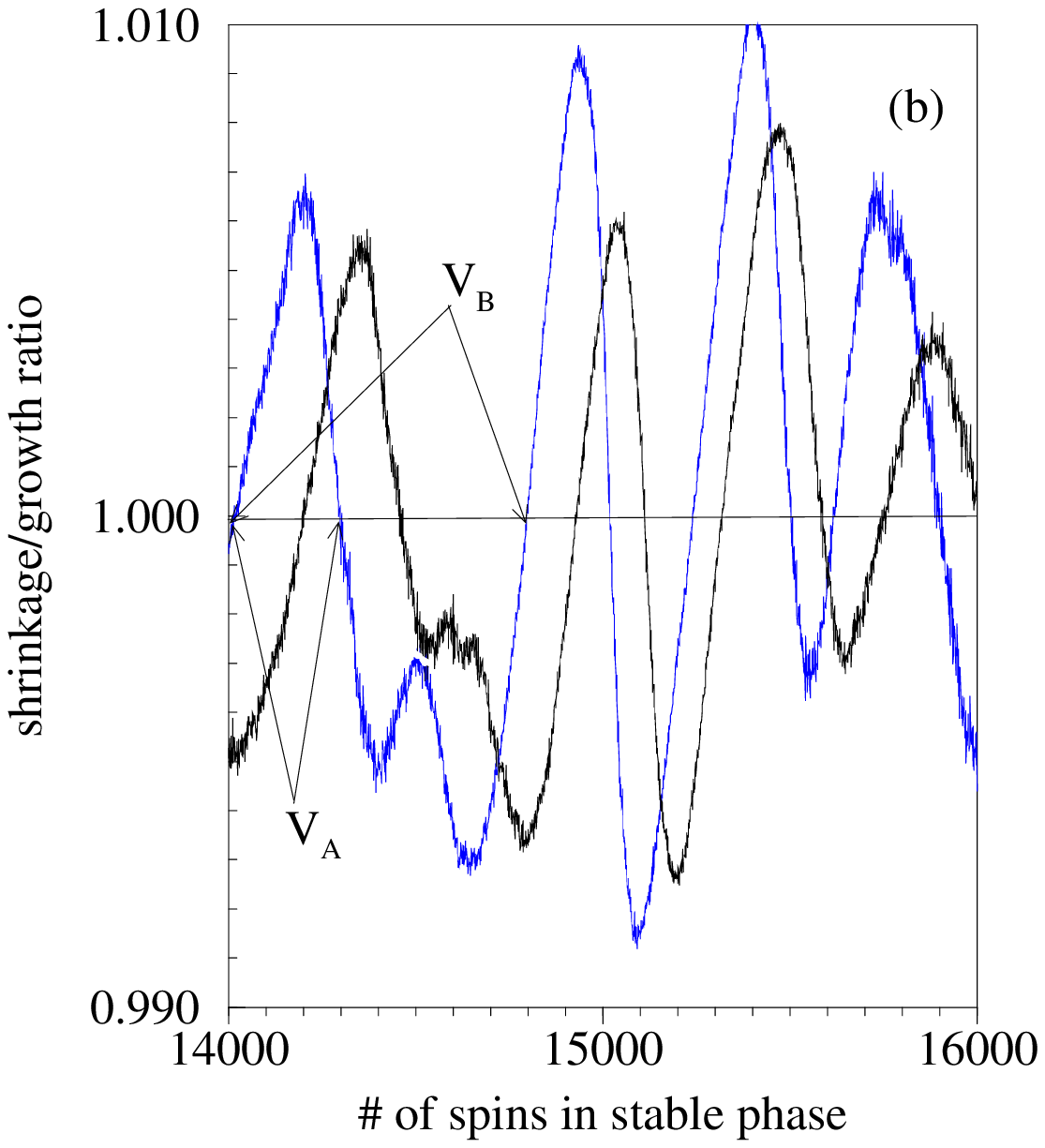}
\caption[]{
(a): 
The average time spent by the domain wall at positions with a given number $n$ of
overturned spins (volume of the stable phase) for several driving
fields. The curves from top to bottom correspond to fields $H=$ 0.03, 0.04, 0.05,
0.06 and 0.07 (in units of $J$).
(b): 
The ratio of the average shrinkage and growth probabilities of the stable phase,
$s(n)/g(n)$.
In regions with values greater than one the domain wall tends to
recede, whereas in regions where this ratio is less than one the domain
wall tends to advance. The curves were obtained for fields
0.03 and 0.04 $J$. The weaker the field, the stronger the ``oscillations''
of the curve. Examples of the activation volume, $V_A$, and of the
corresponding Barkhausen volume, $V_B$, are shown.
}
\end{figure}

Using methods described in detail in Ref.~[\Kolesikb ], one can utilize
$g(n)$ and $s(n)$ to calculate the average time the domain wall
spends in configurations with a given  $n$. 
This function is shown in Fig.~2(a) for several field
strengths. The first thing one notices is that the variation of the
residence time with $n$ increases dramatically  as the field decreases.
While in strong fields the residence times exhibit only small variations,
and the domain-wall velocity is  well defined even on small
length scales, in weak fields the residence-time function exhibits
pronounced peaks. They correspond to the metastable, preferred
configurations of the domain-wall shape. 
Closer inspection reveals that these peaks shift to the right
as the field increases. Thus, the shapes of the favorable
domain-wall configurations depend on the driving force. This is easy
to understand as a ``bulging'' effect: the interface between the pinning centers
is deformed by the field. Therefore, the metastable configurations
contain more of the stable phase in stronger fields.

An important conclusion following from observation of the residence times is that 
the waiting times in individual metastable configurations (the weights of 
the residence-time peaks)
are broadly distributed. 
Another notable observation is that the relative weights of the 
residence-time peaks change with the field.  
This indicates that the pinning effectiveness 
of a given local environment depends on its wider vicinity and perhaps 
on the history of the domain wall (for different values of $H$, the domain wall 
can enter a given region
from different directions, thereby experiencing different
detailed environments).

The distance between neighboring residence-time peaks is the Barkhausen volume.
This can, under suitable conditions, be an experimentally observable
quantity [\Gadetsky ]. However, it must not be confused with the activation volume.
To see the latter, one must examine the growth and shrinkage probabilities.
Figure 2(b) shows the ratio $s(n)/g(n)$ for two different field strengths
in the vicinity of the most pronounced peaks in Fig.~2(a). The domain wall
tends
to retreat from positions in which this ratio exceeds unity, whereas advancing
 is more probable when this ratio is less than one.
The metastable configurations correspond to the points where $s(n)/g(n)$$=$$1$ and
$\partial_{n} [ s(n)/g(n) ]$$>$$0 $. In this figure we clearly see the shifts
in these
positions to the right with increasing field. The distances between these points 
and the subsequent points where 
$s(n)/g(n)$$=$$1$ and $\partial_{n} [ s(n)/g(n) ]$$<$$0$ are the quantities which interest us most. They are the
volumes of
the critical fluctuations, or the activation volumes 
(as indicated by $V_A$ in Fig.~2(b)). 
Like the waiting times, the activation volumes differ widely. More
importantly,
they increase with decreasing field. In other words, the activation volumes
are not simply related to the size of an island or any other volume 
directly defined by the lattice morphology.  Unfortunately, because of the
narrow field region in which our simulations are feasible, it is difficult 
to say what the functional form of the field dependence is [\Kolesik ].

\vspace{0.4truecm}
\noindent CONCLUSIONS 
\vspace{0.4truecm}

Compared to experiment, in computer simulations it is much easier
to look at the processes and quantities which govern the dynamics of 
domain-wall motion in a magnetic medium.
We used the novel approach of Ref. [\Kolesikb ] to identify the thermally
activated fluctuations of domain walls which are responsible for the
domain wall motion, and to measure parameters of these fluctuations.
In particular, we were able to measure Barkhausen volumes and 
activation volumes at different locations in the system. We demonstrated
that in addition to the disorder-induced distribution of activation
volumes, each individual activation volume increases as the strength
of the applied field decreases.
We emphasize that this effect, mostly neglected in the
experimental literature, plays an important role in the field
dependence of the domain-wall velocity.
Although here we only studied a particular model for
iron sesquilayers on tungsten,
our observations are 
expected to apply to other systems in which domain-wall
motion is thermally activated.

\vspace{0.4truecm}
\noindent ACKNOWLEDGMENTS 

This research was supported by NSF Grant No. DMR-9520325,
FSU-MARTECH and FSU-SCRI (DOE Contract No. DE-FC05-85ER25000).

\newpage
\samepage{
\noindent REFERENCES 
\vspace{0.4truecm}
\begin{enumerate}

\item[\Durkop .]
T.\ D\" urkop, H.\ J. Elmers and U. Gradmann,
J.\ Magn.\ Magn.\ Mater. {\bf 172} L1 (1997).

\item[\ElmersA .]
H.\ J. Elmers, J. Hauschild and U. Gradmann,
Phys.\ Rev.\ B {\bf 54},  15224 (1996).

\item[\Elmersb .]
H.\ J. Elmers, J. Hauschild, H. H\" oche, U. Gradmann,
H. Bethge, D. Heuer and U. K\" ohler,
Phys.\ Rev.\ Lett. {\bf 73}, 898 (1994).

\item[\ElmersB .]
H.\ J. Elmers, J.\ Hauschild and U. Gradmann,
J.\ Magn.\ Magn.\ Mater. {\bf 140-144} 1559 (1995).

\item[\Sandera .]
D. Sander, R. Skomski, C. Schmidthals, A. Enders and J. Kirschner,
Phys.\ Rev.\ Lett. {\bf 77}, 2566 (1996).

\item[\Sanderb .]
D. Sander, A. Enders, R. Skomski and J. Kirschner,
IEEE Trans.\ Magn. {\bf 32}, 4570 (1996).

\item[\Skomski .]
R. Skomski, D. Sander, A. Enders and J. Kirschner,
IEEE Trans.\ Magn. {\bf 32}, 4567 (1996).

\item[\Bethge .]
H. Bethge, D. Heuer, Ch. Jensen, K. Resh\" oft and U. K\" ohler,
Surf. Sci. {\bf 331-333}, 878 (1995).

\item[\Elmersa .]
H.\ J. Elmers, J. Hauschild, H. Fritzsche, G. Liu and U. Gradmann,
Phys.\ Rev.\ Lett. {\bf 75}, 2031 (1995).

\item[\Elmersc .]
H.\ J. Elmers and U. Gradmann,
Appl.\ Phys.\ A {\bf 51}, 255 (1990).

\item[\Suen .]
J. H. Suen and J. L. Erskine,
Phys.\ Rev.\ Lett. {\bf 78},  3567 (1997).

\item[\Kolesik .]
M. Kolesik, M.\ A. Novotny and P.\ A. Rikvold
Phys.\ Rev.\ B {\bf 56} 11791 (1997). \\
An animation of the domain-wall motion 
can be found at \\ {\tt http://www.scri.fsu.edu/\~{ }rikvold }

\item[\Pokhil .]
T.\ G. Pokhil and E.\ N. Nikolaev,
IEEE Trans. Magn. {\bf 29}, 2536 (1993).

\item[\Gadetsky .]
S. Gadetsky and M. Mansuripur,
J.\ Appl.\ Phys.\ {\bf 79}, 5667 (1996).

\item[\Kirilyuk .]
A. Kirilyuk, J. Ferr\' e and D. Renard,
IEEE Trans. Magn. {\bf 29}, 2518 (1993).

\item[\Skomskib .]
R. Skomski, J. Giergiel and J. Kirschner,
IEEE Trans. Magn. {\bf 32}, 4576 (1996). 

\item[\Rikvold .]
P.\ A. Rikvold, M.\ A. Novotny, M. Kolesik  and H.\ L. Richards, 
in
{\em Dynamical Properties of Unconventional Magnetic Systems}
edited by A.~T.\ Skjeltorp and D.~Sherrington.
       NATO ASI Series (Kluwer, Dordrecht, in press), cond-mat/9705189

\item[\Kolesikb .]
M. Kolesik, M.\ A. Novotny, P.\ A. Rikvold and D.\ M. Townsley,
in {\em Computer Simulation Studies in Condensed-Matter Physics X},
       edited by D.\ P. Landau, K.\ K. Mon, H.-B. Sch\" uttler 
      (Springer, Berlin, in press), cond-mat/9703115; \\
M.\ Kolesik, M.\ A. Novotny and P.\ A. Rikvold, cond-mat/9710153. 
\end{enumerate}
}

\end{document}